# Electron Spin Resonance investigation of undoped and Li-doped $CdWO_4$ scintillator crystals

V.V. Laguta[1,2], M. Nikl[1], M. Buryi[2]

[1]*Institute of Physics AS CR, Cukrovarnicka 10, 162 53 Prague, Czech Republic*

[2]*Institute for Problems of Material Science, NASc of Ukraine, Krjijanovskogo 3, 03680 Kiev, Ukraine*

**Abstract**

Electron spin resonance (ESR) spectra of $Fe^{3+}$ and $Mn^{2+}$ ions have been studied in the nominally pure and 0.05% Li-doped single crystals of $CdWO_4$. The zero-field splitting parameters are determined with a high precision for both of the impurities: a) $Fe^{3+}$: $b_2^0$ = -7567(10) $cm^{-1}$, $b_2^2$ = 3150(10) $cm^{-1}$; b) $Mn^{2+}$: $b_2^0$ = -1799(10) $cm^{-1}$, $b_2^2$ = 1445(20) $cm^{-1}$. They are in a good agreement with early data for $Mn^{2+}$ (R.E. Donovan and A.A. Vuylsteke, Phys. Rev. **127**, 76 (1962)) but differ by about 5% for $Fe^{3+}$ (Z. Šroubek, K. Ždansky, Czech. J. Phys. B **12**, 784 (1962)). We have found that the doping of $Li^+$ essentially influences the charge balance in the lattice. The result suggest that the Li-doping leads to the increase of the ionic charge of iron from 3+ to 4+ and of manganese, from 1+ to 2+.

## 1. Introduction

Cadmium tungstate ($CdWO_4$) single crystal is widely used intrinsic scintillating material. Its applications mainly focus on the detection of X-rays and γ-rays in medical field [1]. As concerns the density and light yield $CdWO_4$ is superior comparing to other scintillating materials. On the other hand, its practical importance is seriously limited by relatively slow scintillation response. Scintillation characteristics can be degraded by presence of various structural defects and impurities serving as traps or recombination centers for migrating charge carriers generated by ionizing radiation. Usually many of such defects are created by impurity ions which come into the grown crystal from the row materials and crucible. Such defects can be monitored by Electron Spin Resonance (ESR). ESR and related magnetic resonance methods can also provide valuable information on the nature and local structure of trapping centres/defects that is important for further improvement of growth technology.

In the present paper we investigate undoped and 0.05% Li-doped $CdWO_4$ crystals. In particular we have found that the undoped crystal contains mainly only $Fe^{3+}$ ions. On the other hand, Li-doped crystal shows the presence of only $Mn^{2+}$ ions. This suggests that $Li^+$ induces the following charge transformation: $Fe^{3+}$ -> $Fe^{4+}$ and $Mn^+$ -> $Mn^{2+}$. For both $Fe^{3+}$ and $Mn^{2+}$ paramagnetic ions the spectroscopic parameters are also determined too and compared with literature data.

It should be noted that previous ESR studies of $CdWO_4$ were mainly devoted to the characterization of transition metal impurities such as $Mn^{2+}$, $Fe^{3+}$, $Cu^{2+}$, $Mo^{5+}$ (Refs. 2-5) in the crystals which were specially doped by the transition metal ions. Many of the measurements were performed in 60th-70th of the last century with the use of crystals of not too high quality. Therefore early ESR data have to be improved and the spectroscopic parameters have to be determined more accurately.

## 2. Samples and experimental details

The $CdWO_4$ single crystals were grown in air by Czochralski technique using Pt crucibles. The charge was prepared by high temperature solid phase synthesis of stoichiometric composition of cadmium and tungsten oxides. Raw materials with 4N purity were used. Undoped and 0.05% Li-doped $CdWO_4$ crystals were grown. The ESR studies were performed at 9.22 GHz with the standard 3 cm wavelength of the ESR spectrometer; the measurements were carried out in the temperature range 10 – 290 K using an Oxford Instruments cryostat.

$CdWO_4$ crystallizes in a slightly distorted perovskite structure. It is characterized by the space group P2/c [6]. For the ESR measurements, the crystals were oriented, polished and cut in the (100), (010) and (001) planes in a typical shape of about 2x2.5x6 $mm^3$.

## 3. Experimental results

The ESR spectra measured for both the undoped and Li-doped samples are shown in Fig. 1. In the undoped samples there are only two intense resonance lines which change positions under crystal rotation. In order to clarify the origin of this spectrum we carried out the measurements of the spectra in three crystallographic planes. These data are shown in Fig. 2.

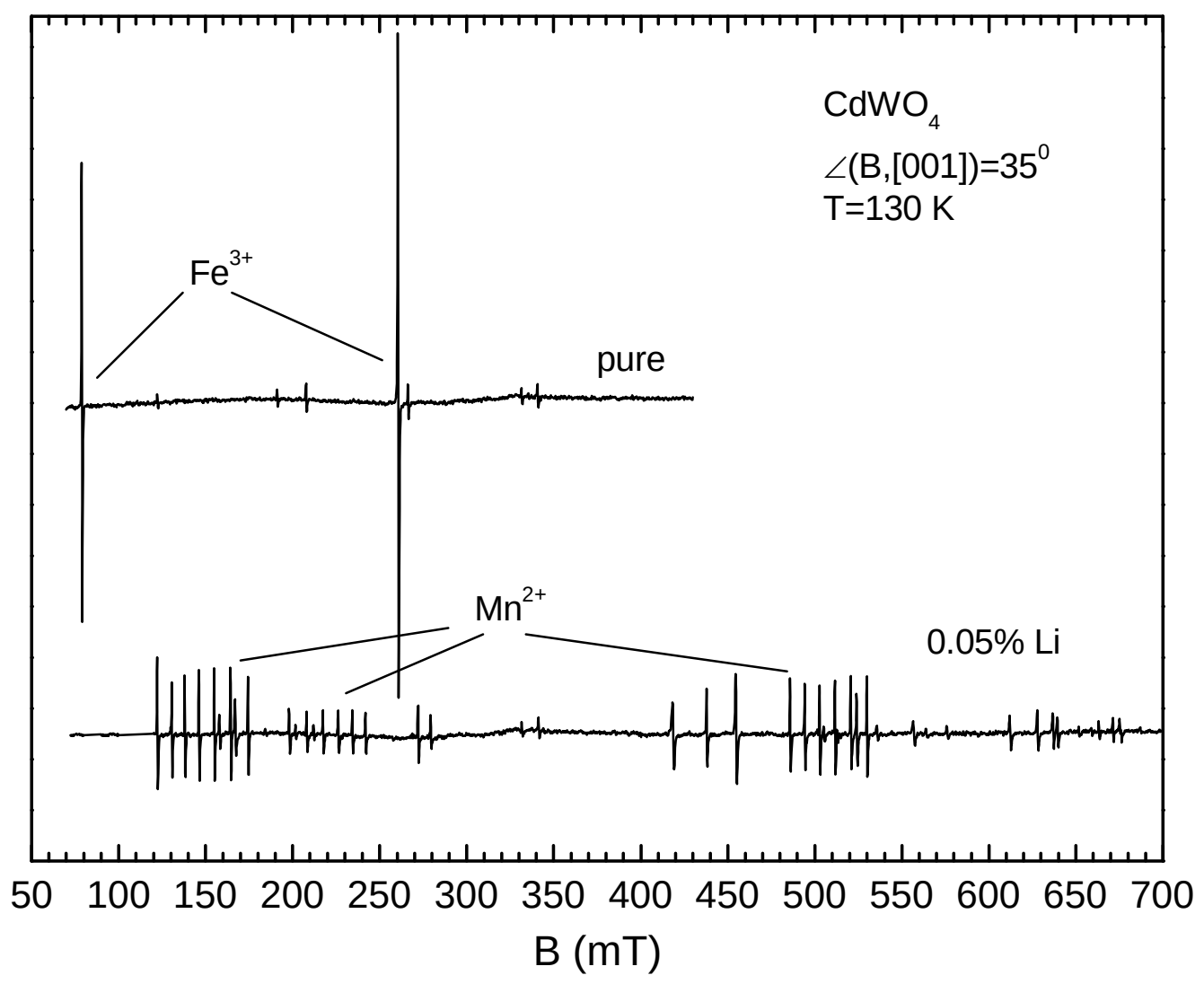

Fig. 1 $Fe^{3+}$ and $Mn^{2+}$ EPR spectra in undoped and Li-doped $CdWO_4$ single crystals measured at 130 K and the magnetic field orientation along the magnetic axis **z**. Hyperfine sextets are well visible for the $Mn^{2+}$ spectrum.

The angular dependencies are very similar to those of $Fe^{3+}$ described previously by Šroubek and Ždansky [3]. Therefore for the description of the resonance fields we have used the spin Hamiltonian of the orthorhombic symmetry for a paramagnetic particle with the spin S=5/2:

$$\widehat{H} = \beta\widehat{\mathbf{S}}\cdot\mathbf{g}\cdot\mathbf{B} + \frac{1}{3}(b_2^0 O_2^0 + b_2^2 O_2^2), \qquad (1)$$

where we included only the second rank terms of the crystal field. Here $b_2^0 = D$ and $b_2^2 = 3E$ are the axial and the rhombic crystal-field constants.

Early study [3] has shown that the Zeeman term in the equation (1) is comparable in value with the crystal-field terms. Therefore the fit of the measured angular dependencies was performed with the numerical diagonalization of the spin Hamiltonian (1). The results are shown in Fig. 2 by the smooth lines.

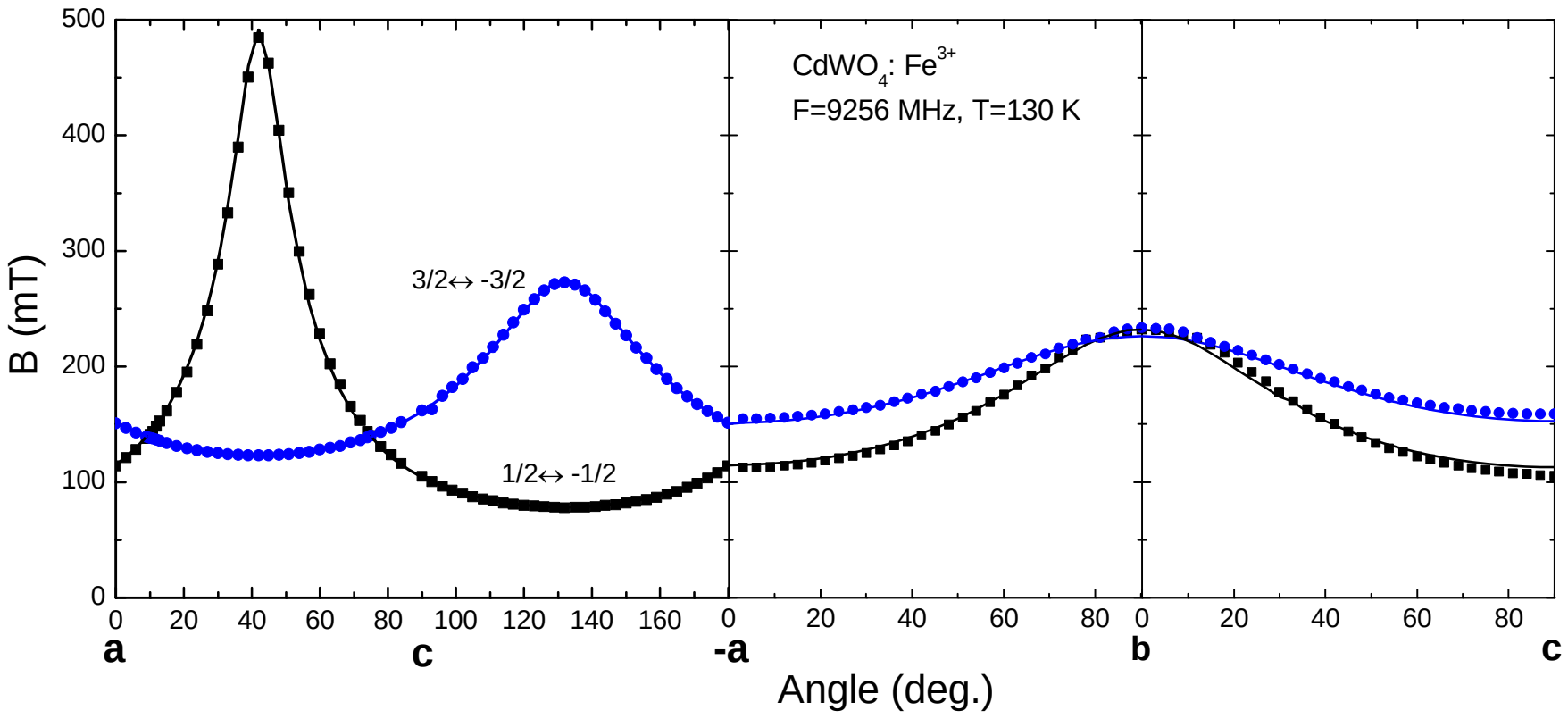

Fig. 2 Angular dependence of $Fe^{2+}$ resonance fields in undoped $CdWO_4$ for the rotation of the magnetic field in the ac, bc and ab planes. Filled squares and solid lines are the measured and the calculated resonances, respectively.

One can see a good agreement between the measured and the calculated angular dependencies for all three crystallographic planes. This indicates that the forth rank terms of the crystal field are small as compared with the second rank terms. The fitting parameters are the follows: g = 2.002(2), $b_2^0$ = -7567(10) $cm^{-1}$, $b_2^2$ = 3150(10) $cm^{-1}$. The signs of the zero-field splitting constants could not be determined unambiguously from our measurements. We can only conclude that $b_2^2$ must be of opposite sign to $b_2^0$. However the sign of the axial constant is known from the low-temperature (T=2K) measurements of $Fe^{3+}$ in another tungstate $MgWO_4$ [7]. It is negative. The magnetic principal axes frame is rotated with respect to the *abc* crystal axes by the angle of 48 deg. in the *ac* plane from the *a* axis towards the *c* axis as it is shown in Fig. 3. The ratio *E/D* = -0.139 is slightly smaller than that determined in early study [3].

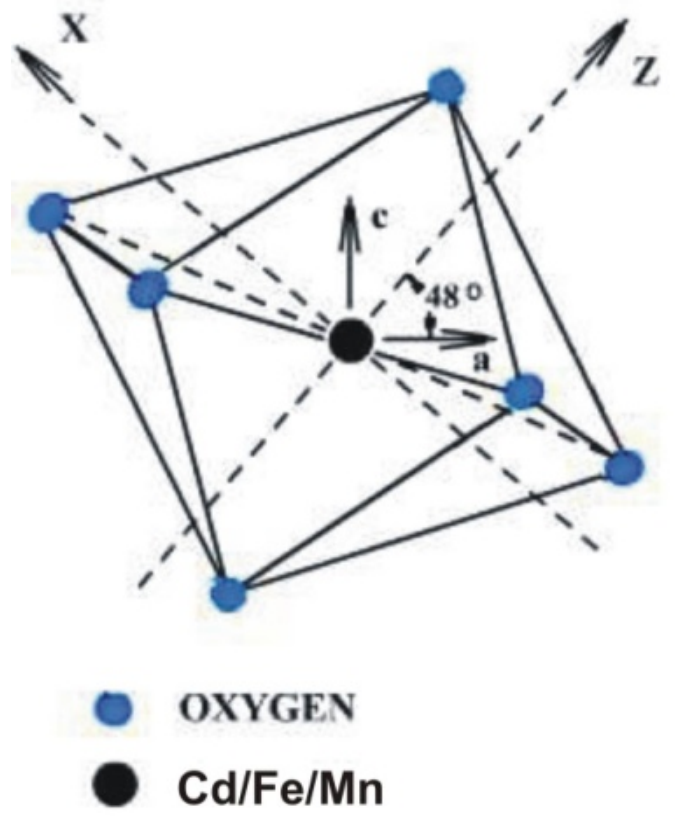

Fig. 3 Positions of oxygens surrounding Fe/Mn ions in $CdWO_4$, projected along $b \equiv y$ axis. Magnetic and crystallographic axes are denoted as *x,y,z* and *a,b,c*, respectively.

In the crystal doped with $Li^{+}$ ions, which are assumed to substitute $Cd^{2+}$ ions due to the small difference of their ionic charges, the $Fe^{3+}$ spectrum completely disappears (Fig. 4). Instead, another spectrum appears which consists of several groups of six-lines each. Such a spectrum is attributed to $Mn^{2+}$ ions ($3d^5$, S=5/2). Due to the nuclear spin of the $^{55}Mn$ isotope (I=5/2, natural abundance 100%) each electron transition is split into six approximately equidistant hyperfine (HF) components. Like in the case of $Fe^{3+}$ impurity ions, we have measured the angular dependencies of the resonance lines in three crystallographic planes at the temperature about 130 K where all resonance lines are well visible. These data are shown in Fig. 4 for the two planes (*ac*) and (*ab*) for the center of gravity of each HF sextet. Consequently in our analysis we did not fit the $^{55}Mn$ HF structure of the $Mn^{2+}$ spectrum. In particular this was related to a complexity of the HF structure as many forbidden HF transitions appear when the magnetic field is turned from the direction of the principal magnetic axes.

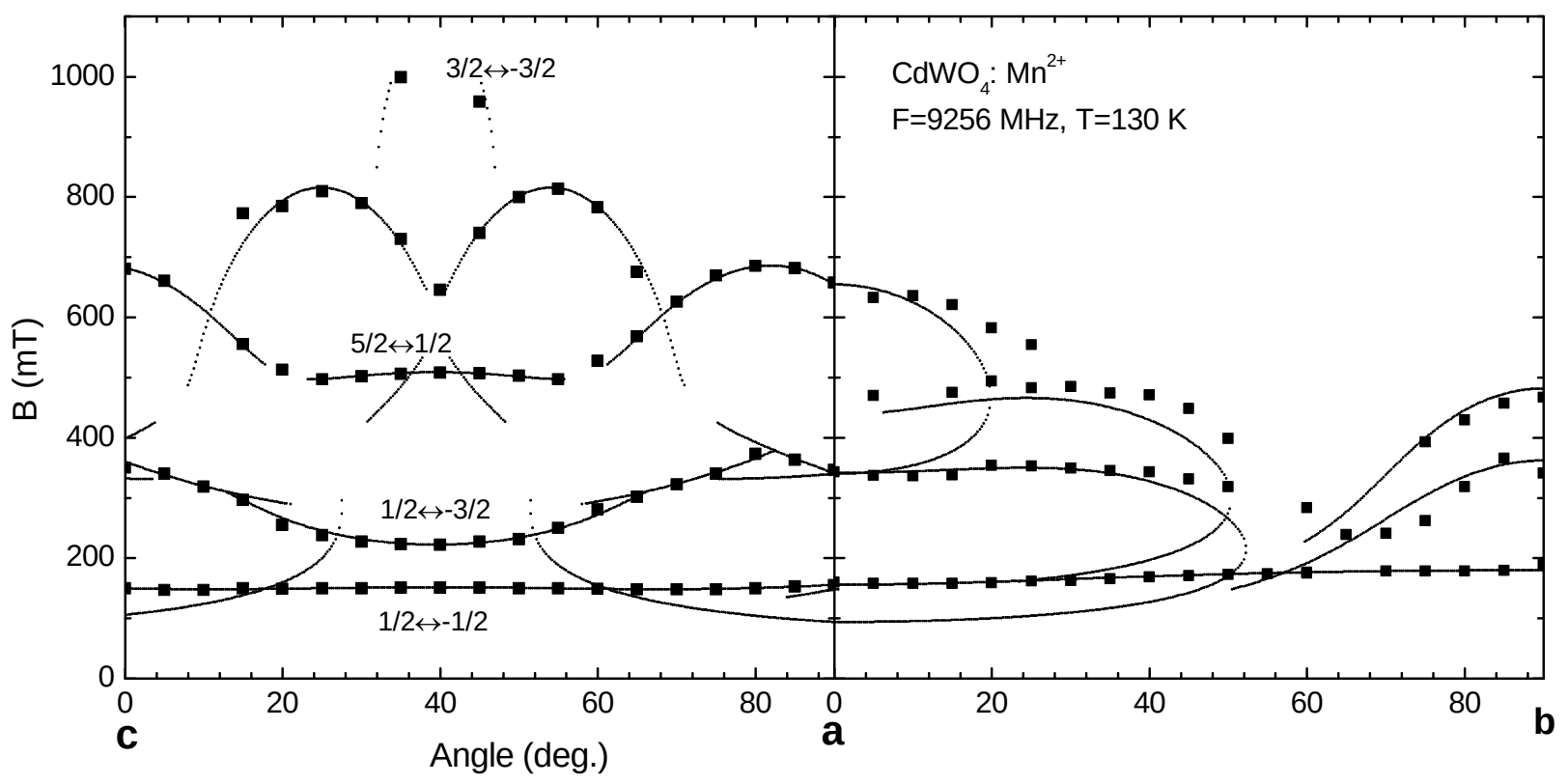

Fig. 4 Angular dependence of $Mn^{2+}$ resonance fields in Li-doped $CdWO_4$ for the rotation of the magnetic field in the ac and ab planes. Filled squares and doted lines are the measured and the calculated resonances, respectively.

$Mn^{2+}$ resonance fields of the electron transitions were described by the spin Hamiltonian (1). The calculated angular dependencies are shown in Fig. 4 by the smooth lines. Note that for $Mn^{2+}$ five electron transitions are visible. However their assignment to different $M_S$ <--> $M_S$ - 1 transitions associated with the $S_Z$ projection of the electron spin is practically impossible as the wave functions are essentially mixed by low-symmetry crystal field terms. The determined spin Hamiltonian parameters are the following: g=2.00(1), $b_2^0$ = -1799(10) $cm^{-1}$, $b_2^2$ = 1445(20) $cm^{-1}$. These parameters are in agreement with those determined in Ref. 2. The magnetic principal axes frame is rotated with respect to the *abc* crystal axes by the angle of 40 deg. in the *ac* plane from the *a* axis towards the *c* axis. The accuracy of the fit of the $Mn^{2+}$ resonances is not as good as in case of $Fe^{3+}$ spectrum. This is mainly connected with the ignoring of the Mn nuclear HF structure. As a result, the experimental resonance fields were determined with large deviations due to the presence of numerous intense forbidden transitions. However even in this case the overall accuracy of the determination of the spin Hamiltonian parameters is relatively high due to the use of the numerical diagonalization of the spin Hamiltonian.

Due to the equality of the $Cd^{2+}$ and $Mn^{2+}$ ionic charges it is expected that Mn impurity sits at the Cd site. This is also supported by the fact that the magnetic axes of both $Mn^{2+}$ and $Fe^{3+}$ as well as of $Cu^{2+}$ [4] have approximately the same orientation with respect to the crystal axes, and these directions agree with the symmetry and distortion of the $CdO_6$ octahedron (see, e.g., Ref. 6).

As both the undoped and Li-doped crystals were grown from the same row materials we assume that the crystals contained both the Fe and Mn impurities. In the undoped crystal Fe was in the 3+ charge state and therefore it was visible in EPR. The charge state of Mn was most probably 1+. In this case the impurities do not change charge balance in the lattice. A $Mn^{+}$ ion being introduced into the lattice has $3d^6$ (S=2, L=2) electronic shell. Due to even electron spin, EPR of the such ion is usually very complicated and needs high MW frequencies [8]. At least, our measurements did not revealed $Mn^{+}$ spectrum. $Li^{+}$ is a very stable ion to the ionization. Therefore it changes the charge balance in the lattice which can be compensated by other impurities and/or intrinsic defects. Fe and Mn, being transition metal ions, can easily change their valence state and thus effectively restore charge balance in the lattice that we indeed observe. Both Fe and Mn increase their valency with the Li doping. Manganese becomes $Mn^{2+}$ and therefore it is easily visible in EPR. Contrary, $Fe^{3+}$ transforms into $Fe^{4+}$ ($3d^4$, S=2, L=2) which again, like $Mn^{+}$, is not favorable for the EPR measurements. Therefore in the Li-doped sample we observe only the $Mn^{2+}$ spectrum. Its integral intensity is approximately equal to that of $Fe^{3+}$ of undoped sample.

In conclusion, we have identified paramagnetic $Fe^{3+}$ and $Mn^{2+}$ ions in undoped and 0.05% Li-doped single crystals of $CdWO_4$. Their ESR spectra were described by the spin Hamiltonian of the orthorhombic symmetry with the following parameters: a) $Fe^{3+}$: g = 2.002(2), $b_2^0$ = -7567(10) $cm^{-1}$, $b_2^2$ = 3150(10) $cm^{-1}$; b) $Mn^{2+}$: g=2.00(1), $b_2^0$ = -1799(10) $cm^{-1}$, $b_2^2$ = 1445(20) $cm^{-1}$. We have found that the doping of $Li^{+}$ essentially changes the ionic charges of both the Fe and Mn impurities. The measurements suggest that with Li-doping $Fe^{3+}$ transforms into $Fe^{4+}$ and $Mn^{+}$ into $Mn^{2+}$.

## Acknowledgements

Financial support of the Czech project GA AV IAA100100810 and the institutional research plan AVOZ10100521 is gratefully acknowledged. The authors are grateful to L.L. Nagornaya for providing $CdWO_4$ crystals.

### References

[1] G. Blasse, B.C. Grabmaier, *Luminescent Materials* (Springer, Berlin, 1994).

[2] R.E. Donovan and A.A. Vuylsteke, Phys. Rev. **127**, 76 (1962).

[3] Z. Šroubek, K. Ždansky, Czech. J. Phys. B **12**, 784 (1962).

[4] Z. Šroubek and K. Ždansky, J. Chem. Phys. **44**, 3078 (1966).

[5] N.Y. Garces, M.M. Chirila, H.J. Murphy, J.W. Foise, E.A. Thomas, C. Wicks, K. Grencewicz, L.E. Halliburton, N.C. Giles, J. Phys. Chem. Solids **64**, 1195 (2003).

[6] P. J. Coing-Boyat, Acta Cryst. **14**, 1100 (1961).

[7] M. Peter, Phys. Rev. **113**, 801 (1959).

[8] A. Abraham and B. Bleaney, *Electron Paramagnetic Resonance of Transition Ions* (Clarendon Press, Oxford, 1970).